\documentclass{article}%
\usepackage[utf8]{inputenc}
\usepackage[english]{babel}
\usepackage{graphicx}
\usepackage{wrapfig,caption}
\usepackage{algpseudocode,tikz}
\usepackage{xcolor}
\usepackage[normalem]{ulem}
\usepackage{authblk}
\usepackage{algorithm}
\usepackage{algpseudocode}
\usepackage{amsmath}
\usepackage{hyperref}
\usepackage{setspace}
\usepackage{lineno}
\usepackage{cleveref}
\usepackage{comment}
\usepackage{bm}
\usepackage[a4paper, total={6.77777in, 9in}]{geometry}
\usepackage[font=footnotesize]{caption}
\usepackage[style=authoryear,date=year,backend=biber,useprefix=true]{biblatex}
\title{Building functional and mechanistic models of cortical computation based on canonical cell type connectivity}

\author[a,b,c,$\ast$]{Arno Granier}
\author[d]{Katharina A Wilmes}
\author[a]{Mihai A Petrovici}
\author[a,e]{Walter Senn}

\affil[a]{{\small Department of Physiology, University of Bern, Switzerland}}
\affil[b]{{\small Graduate School for Cellular and Biomedical Sciences, University of Bern, Switzerland}}
\affil[c]{{\small Group for Neural Theory, Ecole Normale Supérieure, France}}
\affil[d]{{\small Institute of Neuroinformatics, University of Zurich and ETH Zurich, Zürich, Switzerland}}
\affil[e]{{\small Center for Artificial Intelligence in Medicine, University of Bern, Switzerland}}
\affil[$\ast$]{{\small correspondence: arno.granier@ens.psl.eu}}
\begin{document}
\maketitle

\begin{abstract}\small
  \noindent Neuronal circuits of the cerebral cortex are the structural basis of mammalian cognition. The same qualitative components and connectivity motifs are repeated across functionally specialized cortical areas and mammalian species, suggesting a single underlying algorithmic motif. Here, we propose a perspective on current knowledge of the cortical structure, from which we extract two core principles for computational modeling. The first principle is that cortical cell types fulfill distinct computational roles. The second principle is that cortical connectivity can be efficiently characterized by only a few canonical blueprints of connectivity between cell types. Starting with these two foundational principles, we outline a general framework for building functional and mechanistic models of cortical circuits. Finally, we show that the compressed and generative representations of connectivity entailed by this framework are particularly amenable to automatic search through gradient-free optimization, providing a concrete first step towards modelling the evolution of functional neural circuits.
\end{abstract}

\section{Introduction}
Recent technical advances have enabled a substantial increase in comprehensive data regarding neuronal circuits of the cerebral neocortex (hereafter simply `cortex'). Transforming this emerging data into a mechanistic understanding of computation in cortical circuits requires theoretical and computational perspectives that are both informed and constrained by the data.

Core constituents and connectivity motifs of the cortical architecture are largely conserved across mammalian species and cortical areas \parencite{Hubel1974-rv,Mountcastle1997-mb,Douglas2004-im,Sur2005-zs,Kaschube2010-hb,Harris2015-fi,Powell2024-tp,Meyer2025-te} despite some area-specific and species-specific specializations \parencite{Tasic2018-yf,Hodge2019-ez,Bakken2021-nz,Fang2022-ni,Loomba2022-jf,Ma2022-xc}. The relative uniformity of the cortex is of interest for theoreticians, as it potentially underlies a single algorithmic motif central to cortical computation.

In this work, we present a computational perspective on current knowledge of cortical structure. We do not strive for a detailed account of biophysical properties \parencite{Billeh2020-ko,Dura-Bernal2023-at,arkhipov2025integrating}, and rather aim to provide a description at a level of abstraction that is both grounded in biology and a necessary component for understanding cortical computation at the mechanistic level \parencite{Bernaez-Timon2023-we}. We conclude that different cortical cell types map to different functions and computations, and that cortical connectivity can be efficiently recapitulated through canonical connectivity blueprints between cell types. On the premise of these two core principles extracted from the literature, we outline a general framework for constructing models of cortical computation centered on the notion of canonical connectivity among cortical cell types.

\section{Cortical cell types are functionally specialized}

\begin{figure}[!ht]
  \includegraphics{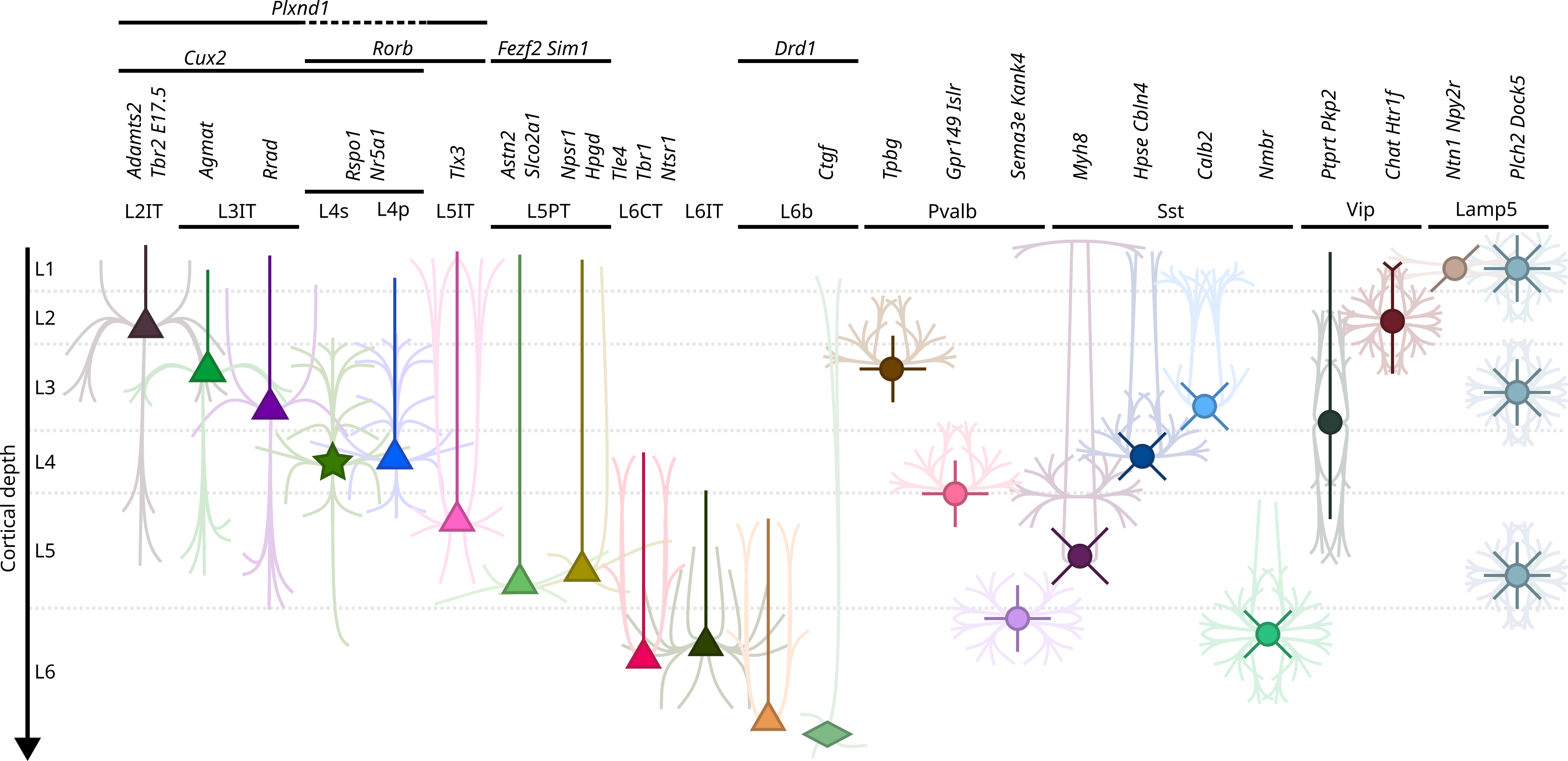}
  \caption{
    Cortical neuron types. The extent of the apical dendritic trees is represented in dark colors, and patterns of local axonal projections in light colors. As sketches, these have some degree of arbitrariness and do not represent the variability of arborization within one class. Marker genes defining a cell type are indicated in italics. Excitatory cells can be named classically based on their laminar position (L2IT - L6b; IT: intratelencephalic, s: stellate cells, p: pyramidal, PT: pyramidal tract, CT: corticothalamic) and inhibitory cells based on expressed molecular markers (Pvalb, Sst, VIP, Lamp5). Excitatory cells are pyramidal in shape, except for L4s and L6b which have different morphologies. Interneuron subtypes depicted here are representative subtypes from the Allen Cell Types Database, notably as reported by \textcite{Patino2024-ts} and \textcite{Wu2023-jm}. Pyramidal subtypes depicted are the major classes while more minor classes, such as near-projecting layer 5 pyramidal cells and Car3 expressing layer 6 pyramidal cells, are not represented individually. Good entry points for a catalog of those genes are the work of \textcite{Matho2021-cs} or the Allen Cell Types Database. Note that gene expression varies greatly across areas for excitatory but not inhibitory cells \parencite{Tasic2018-yf}. Consequently, some genes presented here are area-specific, such as Adamts2, Agmat, and Rrad, which characterize layer 2/3 in primary visual cortex but not in anterior lateral motor cortex \parencite{Tasic2018-yf}.} \label{fig:types}
\end{figure}

A first step towards understanding the functional architecture of the cortex is to map its basic constitutive elements: cortical cell types \parencite{Zeng2022-vw}, see \cref{fig:types}. Excitatory (glutamatergic) neuron types, typically pyramidal in shape, have primarily been characterized by their laminar position, morphology, and the brain structures targeted by their axons, while inhibitory (GABAergic) interneuron types have been characterized by key molecular markers, morphology, and patterns of local connectivity. In addition to these classical accounts, recent efforts to develop precise and complete taxonomies focus on the unbiased definition of cell types as clusters of cells with similar gene expression profiles \parencite{Brain_Initiative_Cell_Census_Network2021-cg,Chen2023-ne,Siletti2023-ko,Yao2023-el}. Characterizations based on genetic, anatomy, or connectivity yield a consistent picture of the diversity of broad cell types \parencite{Gouwens2020-ii,Yuste2020-hq,Peng2021-xu,Scala2021-ey,Zhang2021-xw}, resulting from precisely orchestrated developmental mechanisms \parencite{Lim2018-uu,Huilgol2023-sh,Di_Bella2024-yz}.

The functional specialization of broad interneuron classes defined by key molecular markers (Pvalb, Sst, VIP) is well established \parencite{Tremblay2016-wy}. Recent work on inhibition in layer 1 completes the picture by showing that Ndnf/Lamp5 cells compete with Sst cells for the dendritic inhibition of pyramidal cells \parencite{Abs2018-hm,Pardi2023-bx,Naumann2025-te}. A subdivision of these broad classes into finer cell types appear functionally warranted since, beyond the high degree of cell-type-specificity of interneuron connectivity (see next section), emerging evidence indicates that finer interneuron types within these broad classes also show distinct functional responses \parencite{Munoz2017-rh,Bugeon2022-ib,Green2023-vy}. Notably, \textcite{Green2023-vy} report that one specific cell type within the Sst class in the mouse posterior parietal cortex carries a strong error-correction signal. 

Different excitatory cell types also show clear differences in functional response patterns, likely underlying specialized computational roles. For example, pyramidal cells of cortical layer 5 that project subcortically (pyramidal tract, PT) and those that do not (intratelencephalic, IT) exhibit distinct activity profiles during decision-making \parencite{Musall2023-oc}, motor control \parencite{Mohan2023-kv}, sensory detection \parencite{Takahashi2020-pd}, and associative learning \parencite{Moberg2025-qi}. \textcite{vo2025typespyramidalcellsrole} suggest a general account of these differences based on the reported unidirectional connectivity from IT to PT, whereby the activity of PT pyramidal cells directly depends on the activity of IT pyramidal cells, but not conversely. For instance, in the context of reinforcement learning, \textcite{Moberg2025-qi} propose that this unidirectional connectivity supports the computation of an estimate of the stimulus value (in PT) using its sensory representation (in IT). Other examples of the functional specialization of excitatory cell types include reports that pyramidal types that project to different subcortical regions assume different roles in motor control \parencite{Economo2018-mw} and the regulation of social competition \parencite{Xin2025-jd}, and that pyramidal types of cortical layer 2 and layer 3 show distinct response patterns to sensory stimuli \parencite{Condylis2022-iy} and prediction errors \parencite{OToole2023-qc}. Finally, pyramidal types are differentially impacted by anesthetics \parencite{Bharioke2022-ks} and antipsychotic drugs \parencite{Heindorf2024-aq}, with the effects observed specifically on pyramidal cells of cortical layer 5.

In summary, recent studies emphasize the fine-grained functional specialization of cortical cell types, likely underlying their distinct computational roles in the cortical algorithm.
A key determinant of these differences is the cell-type-specific nature of cortical connectivity, reviewed next.

\section{The structure of local cortical connectivity} \label{sec:local}

In this section, we focus on the patterns of local connectivity between different cortical cell types. A promising direction for elucidating rules of local connectivity lies in the detailed reconstruction of neural circuits at nanoscale precision using electron microscopy. Recent contributions include reconstructions of (i) a complete `column' of mouse primary somatosensory cortex \parencite{Sievers2024-nv}, of (ii) a cubic millimeter of mouse visual cortex spanning multiple cortical areas \parencite{microns2025functional}, and of (iii) a cubic millimeter of human temporal cortex \parencite{Shapson-Coe2024-ah}.

\begin{figure}[!ht]
  \includegraphics{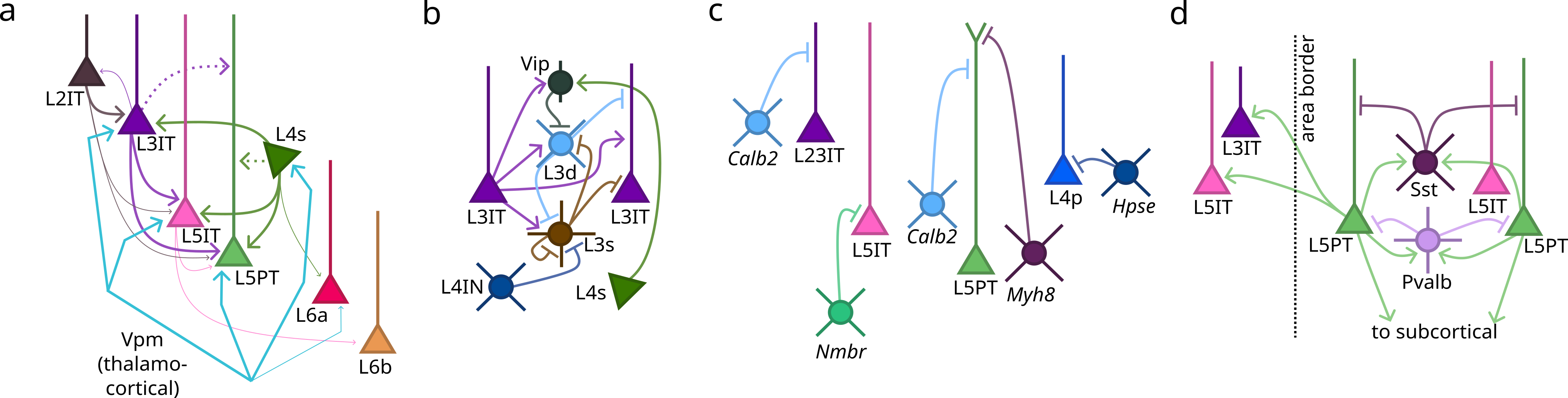}
  \caption{Canonical local cortical circuits. (a) The main information flow in a cortical column of the mouse barrel cortex as reported by \textcite{Sievers2024-nv}. The thickness of the lines indicates the strength of the projection. Dotted lines indicate the dendritic location of projections from L3IT and L4IT to L5PT. Recurrent projections from cells of one type to other cells of the same type are present throughout (not drawn). (b) The logic of intercolumnar feedforward inhibition as reported by \textcite{Sievers2024-nv}. One population of dendrite-targeting (L3d) and one population of soma-targeting (L3s) interneurons per excitatory class/layer mediate feedforward inhibition. These are controlled by local populations via direct connections and disinhibitory connections relaying through a population of bipolar cells (VIP). L4IN: layer 4 interneurons. (c) Cell-type-specific targeting of Sst interneurons subtypes as reported by \textcite{Wu2023-jm}. Expressed genes are reported in italics. (d) Local outputs of L5PT (morphologically-defined thick-tufted layer 5 pyramidal cells) as reported by \textcite{Bodor2023-yt}.
    } \label{fig:local}
\end{figure}

Of particular interest, \textcite{Sievers2024-nv} report a complete matrix of intercolumnar connectivity between cell types (see also \cite{Binzegger2004-ge}), of which the excitatory-to-excitatory part is represented in \cref{fig:local}a. They confirm the strong recurrent connections between excitatory cells of the same type \parencite{Douglas1995-zr}, paralleled by a canonical inhibitory-disinhibitory circuit repeated in each cortical layer and driven by local excitatory cells, see \cref{fig:local}b. A classical view of information processing by different local pyramidal types is as a highly sequential process, from input layer 4 to layer 3 then from layer 3 to output layer 5 \parencite{Douglas2004-im}. This account is completed and challenged by the identification of strong direct pathways from layer 4 to layer 5 \parencite{Sievers2024-nv} and from layer 5 IT to layers 2 and 3 \parencite{Hage2022-ct}, and the finding that direct projections from the thalamus can directly and strongly activate pyramidal cells of layer 5 without relaying through layer 4 \parencite{Constantinople2013-cs}. The structure of local cortical connectivity suggests parallel and recurrent, yet specific information processing pathways.
Finally, \textcite{Sievers2024-nv} confirm the differential subcellular targeting from different excitatory cell types onto pyramidal cells of layer 5 \parencite{Petreanu2009-qw}. 

The classical picture of interneuron circuitry, whereby Pvalb cells target somata and proximal dendrites of nearby pyramidal cells and other Pvalb cells, Sst cells target distal dendrites of all nearby non-Sst cells, and VIP cells target nearby inhibitory cells \parencite{Tremblay2016-wy,Campagnola2022-ps}, is completed by emerging data showing that finer types within those broad classes show distinct cell-type-specific connectivity patterns \parencite{Patino2024-ts,schneider2025inhibitory}. A prime example can be found in Sst subtypes, where \textcite{Wu2023-jm} and \textcite{gamlin2025connectomics} demonstrate highly cell-type-specific connectivity patterns in terms of postsynaptic cell type and targeted subcellular compartment, see \cref{fig:local}c. \textcite{Bodor2023-yt} further show that proximal projections from subcortically-projecting pyramidal cells of layer 5 preferentially target specific interneuron types that in turn preferentially target back the same type of pyramidal cells, see \cref{fig:local}d.

Both for excitatory and inhibitory projections, it is clear that the spatial proximity of axons and dendrites is not sufficient to predict connectivity; rather, cell-type-specific connectivity is the norm, supported by the differential expression of synaptic molecules depending on pre- and post-synaptic cell types \parencite{Favuzzi2019-th,Bernard2022-qg}.
This highlights again the cell type as a central unit for deciphering cortical connectivity and computation.
Moreover, this points to the important level of description of local cortical connectivity as a canonical blueprint of connectivity between cortical cell types.

\section{The structure of interareal connectivity}

In this section, we come to a mesoscale description of the cortex as a network of cortical areas. The developmental and evolutionary processes underlying cortical arealization have been extensively investigated and are reviewed elsewhere \parencite{Cadwell2019-wx,Chen2024-fm}.

\begin{figure}[!ht]
  \centering
  \includegraphics{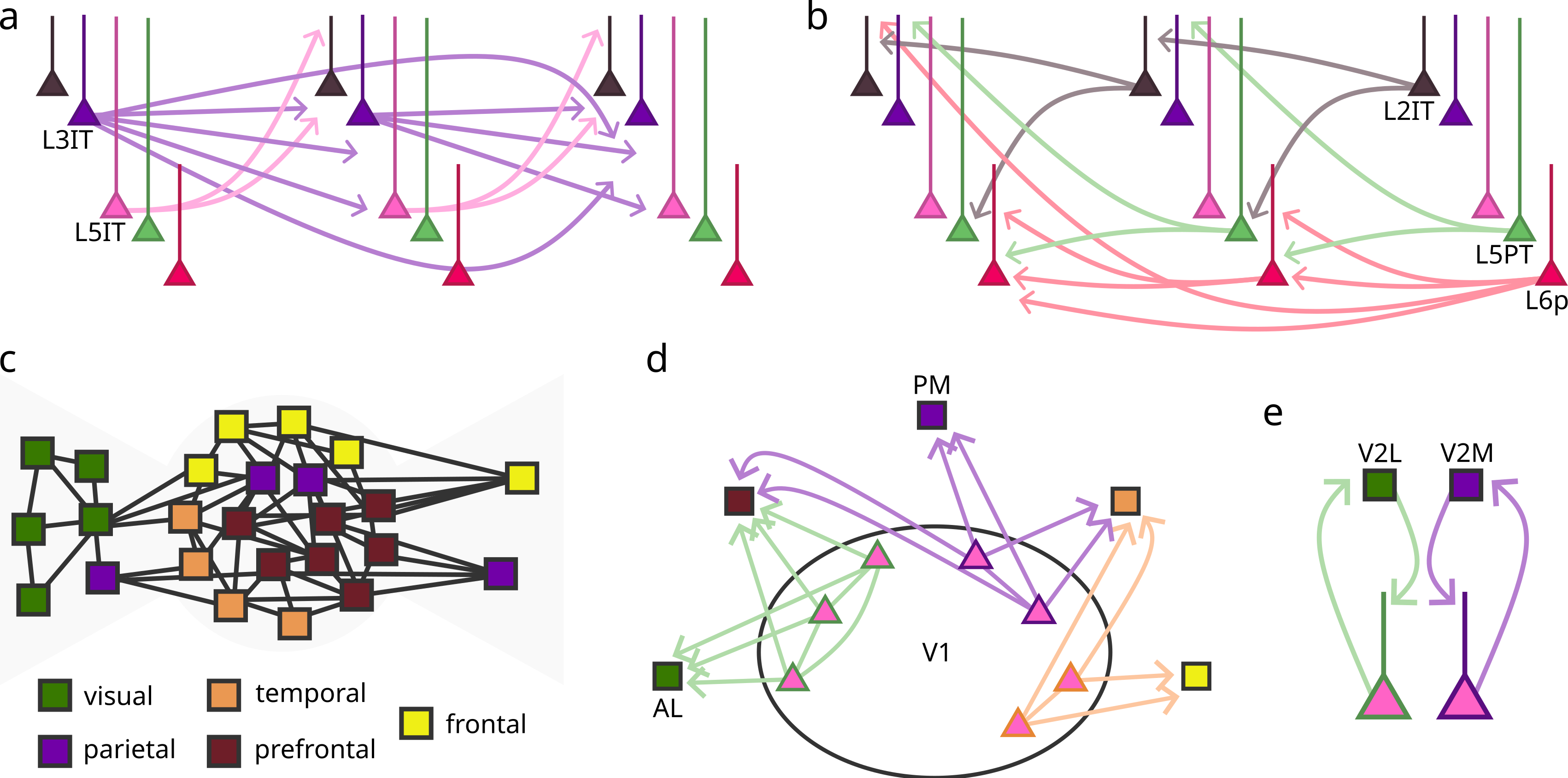}
  \caption{Specific structure of interareal projections. (a,b) Cell types of origin and laminar target patterns of main interareal feedforward (a) and feedback (b) projections, based on integrated data from \textcite{Markov2014-zs} and \textcite{Harris2019-ga}. Note that this mixes data from monkeys and mice. \textcite{Harris2019-ga} additionally report interareal projections from pyramidal cells of layer 4, not represented here. The separation of pyramidal cells from layers 2 and 3 stems from \textcite{Markov2014-zs} (in primates), while \textcite{Harris2019-ga} (in mouse) represent both of these as a homogeneous class (\textit{Cux2}), rendering their laminar target patterns unclear. In both studies, the granularity of layer 6 pyramidal cells could not be accounted for. Finally, \textcite{Harris2019-ga} report that interareal projections from layer 5 PT and layer 6 CT pyramidal cells are less numerous than from IT pyramidal cells. More data is needed on these matters. (c) A sketch of a typical interareal graph of connectivity, with only the strongest connections represented. Each square is an area, and colors represent broad groupings of areas, typically lobes. The spatial arrangement of squares does not reflect the physical placement of areas but is based solely on connectivity (e.g.~following a force-directed graph drawing algorithm). This type of graph but with real data is presented e.g.~by \textcite{Gamanut2018-tl}, their fig.~8. (d) Single pyramidal cells in mouse primary visual cortex project to non-random sets of other areas \parencite{Han2018-xt}. Notably, pyramidal cells in the primary visual cortex projecting to anterolateral (AL) and posteromedial (PM) visual areas are segregated populations with little local interactions \parencite{Kim2018-rk}. (e) Pyramidal cells preferentially receive feedback from areas that they themselves project to, in particular, for pyramidal cells of layer 5, at their apical dendrites \parencite{Kim2020-vl,Siu2021-bg,Young2021-vl}. V2L: lateral visual areas, V2M: medial visual areas.} \label{fig:inter}
\end{figure}

Only a subset of cortical cell types send interareal projections, the majority of which are pyramidal cells. Connections between cortical areas can be characterized on the basis of which specific cell types send and receive projections in the source and target areas respectively (as well as the specific subcellular compartment targeted by the axons).
Similar to local connectivity (see \cref{sec:local}), interareal connectivity can be efficiently characterized by only a few blueprints of connectivity between cortical cell types.
For example, the notion of a cortical hierarchy can be constructed by assigning each interareal connection to one of only two classes (or a continuum between them), traditionally deemed feedforward and feedback \parencite{Rockland1979-ye,Felleman1991-if,Markov2014-zs,Harris2019-ga,Yao2023-bp}. Pyramidal cells sending feedforward and feedback projections form two largely distinct, segregated populations. In primates, \textcite{Markov2014-zs} report that feedforward projections originate from pyramidal cells of layers 3 and 5, while feedback projections originate from pyramidal cells of layers 2 and 6, see \cref{fig:inter}ab. In the target area, \textcite{Harris2019-ga} provide a precise account of cortical layers targeted by long-range projections originating from different cell types, spanning the whole mouse cortex, see \cref{fig:inter}ab; their general account is corroborated by single-cell projectome analysis \parencite{Gao2022-mu}. 

Interareal projections originating from specific cell types might form largely segregated streams of computation across the cortex, of which a classical example is the idea of independent `dual counterstreams' in superficial and deep layers \parencite{Markov2013-iz,Barzegaran2022-na}. Reciprocal interareal connections have been characterized as conveying predictions or expectations and errors respectively. However, it remains unclear how local circuits process predictions, particularly whether or how they explicitly or implicitly compute differences from local activity or other afferent signals \parencite{de-Lange2018-xr,Keller2018-qi,Marques2018-kj,Sacramento2018-pp,Whittington2019-jv,Jordan2020-vc,Lillicrap2020-du,payeur2021burst,Garner2022-sb,greedy2022single,Hertag2022-va,Audette2023-ek,Aceituno2024-tj,Dias2024-qe,Ellenberger2024-iv,Furutachi2024-qj,Seignette2024-oo}. A clearer picture of the functional role of feedback connections will certainly come from their direct manipulations in awake primates \parencite{Debes2023-gl,Andrei2025-ia}, complemented by computational perspectives \parencite{Tugsbayar2024.10.01.615270}.

The graph of connectivity between areas forms clusters of highly interconnected, functionally related, and spatially close areas, with still substantial inter-cluster connectivity \parencite{Markov2013-iz,Zingg2014-st,Gamanut2018-tl}. More precisely, \textcite{Markov2013-iz} characterize a core-periphery organization in the primate cortex, with a densely interconnected core comprising a majority of the cortical areas, see \cref{fig:inter}c. When introducing the notion of feedforward and feedback connections, the resulting directed graph exhibits both hierarchical and nonhierarchical connectivity motifs, and we caution against considering the cortex as a highly hierarchical system. In fact, on a spectrum ranging from random (0\%, every interareal connection is assigned feedforward or feedback randomly) to perfectly hierarchical (100\%, every interareal connection is assigned feedforward or feedback based on the final hierarchical position of the source and target areas), \textcite{Harris2019-ga} place the mouse cortex at 19\%. Still, two areas will often be connected reciprocally (bidirectionally) and asymmetrically (one direction is feedforward and the other feedback). This notion of reciprocal asymmetric connectivity seems to us better equipped to deal with realities of the cortex compared to the notion of a hierarchy.

Finally, we highlight the projection-target heterogeneity of single pyramidal types. Within one genetically defined cell-type-specific population in one area of mouse cortex, different subpopulations project to different non-random subsets of target areas \parencite{Han2018-xt,Kim2018-rk,Munoz-Castaneda2021-xr,Peng2021-xu}, see \cref{fig:inter}d. For example, \textcite{Kim2018-rk} demonstrate that pyramidal cells in superficial layers of primary visual cortex that project to anterolateral cortex or posteromedial cortex constitute segregated populations with limited local interactions. These subpopulations form genetically separable subpopulations within a broader genetically continuous cell type \parencite{Kim2020-vl,Klingler2021-ix}. In turn, subpopulations receive feedback projections preferentially from areas \parencite{Kim2020-vl,Siu2021-bg,Young2021-vl}, streams \parencite{Federer2021-xe}, or even columns \parencite{Watakabe2023-lx} that they themselves project to, see \cref{fig:inter}e. The extent to which this tight reciprocity of feedback is universal remains unclear.

\section{A general framework for functional and mechanistic models of cortical circuits}

Our brief review of cortical structure outlines structural properties that a mechanistic model of cortical computation should accommodate. Here, we sketch a general framework for constructing such models. 

\begin{figure}[!ht]
  \includegraphics{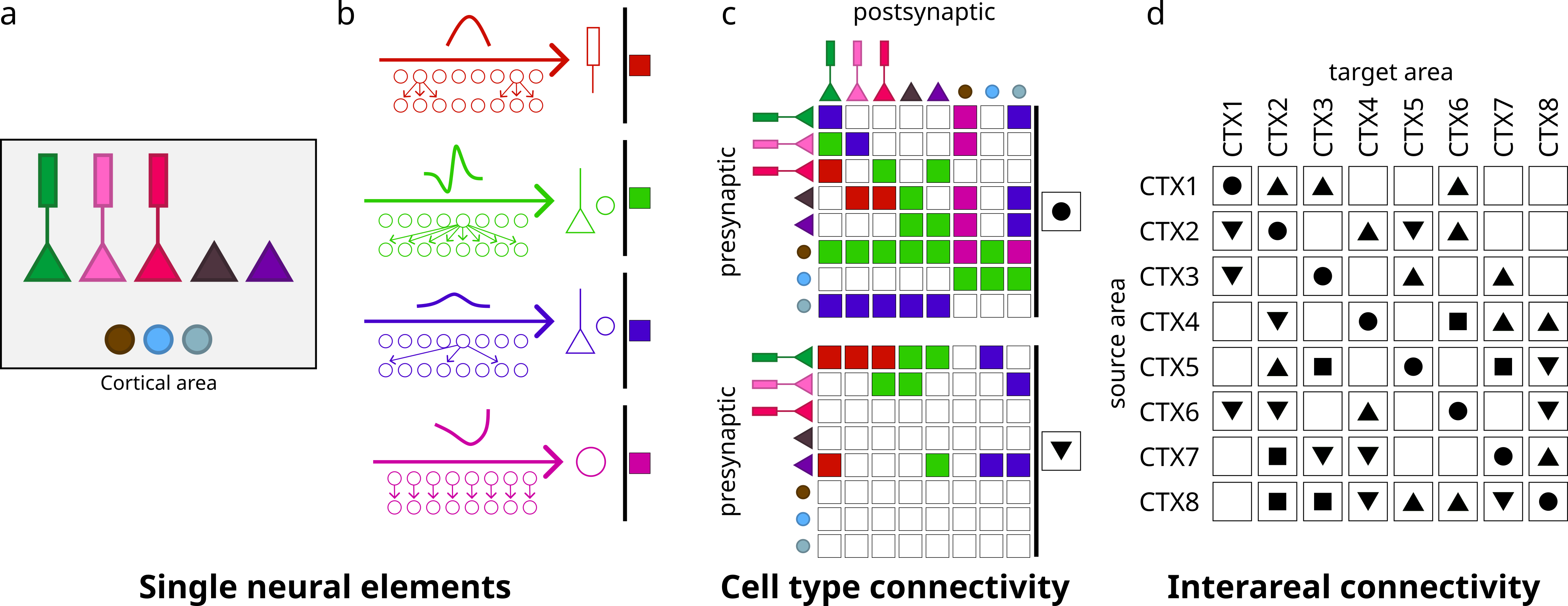}
  \caption{What do we need for a functional and mechanistic model of cortical computation? This figure sketches model elements: the specifics are simply illustrative. (a) Defining cortical cell types. Different shapes of neuron sketches represent different neuron models (details pending), while different colors represent different cell types. Additionally to cortical cell types, we also sketch a thalamic cell type. (b) Defining prototypical synapse types. On top of the arrows we sketch the plasticity kernels, on the right the possible subcellular targets, and on the bottom the connectivity patterns. (c) Sketches of cell type connectivity blueprints matrices. Colors refer to the colors of synapse types as in b. (d) A sketch of an interareal connectivity matrix. Pictograms represent different connectivity blueprints.} \label{fig:general}
\end{figure}

A first step is to define single neuron models, characterized by a set of variables, a parameter space, and a function describing neuronal dynamics. On the basis of one neuron model (shapes in \cref{fig:general}a), multiple cell types can be defined (colours in \cref{fig:general}a). For example, pyramidal cells of all types might be modelled with a two compartment leaky integrate-and-fire model, potentially with parameters varying across types. By starting with the definition of cortical cell types, we emphasize their diversity and functional relevance. The following step is the definition of several prototypical projection types, see \cref{fig:general}b. A projection type is defined by a parameter space, a function describing its dynamics, and a function describing parameter learning. The dynamics of a projection type typically describes how the variables of the postsynaptic neurons change as a function of the variables of the presynaptic neurons and the projection parameters. In defining projection types, it is common to decouple the definition of a single synapse (plasticity kernels, subcellular targets, etc.) and the definition of rules of connectivity at the level of the population (connectivity patterns, distance rules, connection probabilities, etc.). The frame for definitions of neural elements is intentionally kept general. For example, we are agnostic on whether the neuron models use a spike code or a rate code.

After defining cell and projection types, it remains to define connectivity, that is, which cells are connected and with which type of projections. This is the heart of the matter and can be done in two steps. First, define cell type connectivity blueprints as matrices where the row index represents presynaptic cell type, the column index represents postsynaptic cell type, and the value represents synapse type, see \cref{fig:general}c. The simplest would be to have three blueprints : local, feedforward, and feedback; in reality, it is likely more complex (see \cite{Shen2022-uz,Liu2024-ik} for the diversity of feedback target patterns). Second, construct the matrix of interareal connectivity where the row index represents the source area, the column index represents the target area, and the value represents the connectivity blueprint that this connection follows, see \cref{fig:general}d. These matrices are not to be confused with typical weight matrices: their role is to define the instantiated algorithm. The mapping of the structures of these matrices through experiments has been the central topic of the previous sections. Accounting for the fine-grained projection-specificity of cortical cell types (see previous section; \cite{Han2018-xt,Kim2018-rk}) necessitates additionally finer subdivisions of either areas or cell types, but does not change the fundamental procedure.

When designing a particular model of this form, it is useful to refer not only to structural but also functional studies, in particular when choosing which projection model should be applied to the connections from and to specific cell types. For example, it appears that pyramidal cells of the same type (at least in superficial layers) preferentially connect locally based on their selectivity and whether they encode correlated features or are often coactive (\cite{Hopfield1982-ay,Ko2011-iq,Cossell2015-dq,Lee2016-rn,Iacaruso2017-ts,Carrillo-Reid2019-zh,Marshel2019-kr,Miehl2023-vg,Fu2024-of,Oldenburg2024-tr,Yuste2024-ao,ding2025functional}; see also \cite{Chavane2022-qm}). In the local connectivity blueprint, the projection model applied from pyramidal cells of a type to pyramidal cells of the same type (on the diagonal) should reflect this fact, either directly or in the connectivity patterns resulting from its synaptic learning rule (e.g., Hebbian). Similar arguments can be made for the functional organization of projections between pyramidal cells and specific interneuron classes \parencite{Znamenskiy2024-kx,Kuan2024-ul}. Finally, feedback projections from deep pyramidal cells to primary visual cortex follow an orthogonal organization, preferentially overspreading perpendicular to their preferred orientation and opposite to their preferred direction of motion \parencite{Marques2018-kj,Dias2024-qe}. This would be reflected in the row of the feedback blueprint corresponding to deep pyramidal cells.

To simulate the model, a simulation procedure needs to be defined. This simulation procedure tells when and in what order the functions defining neuronal dynamics, projection dynamics, and learning should be activated. A simple, yet biologically plausible solution, is to activate a step of all dynamics and learning functions in parallel at every timestep.

\begin{figure}[!ht]
  \centering
  \includegraphics{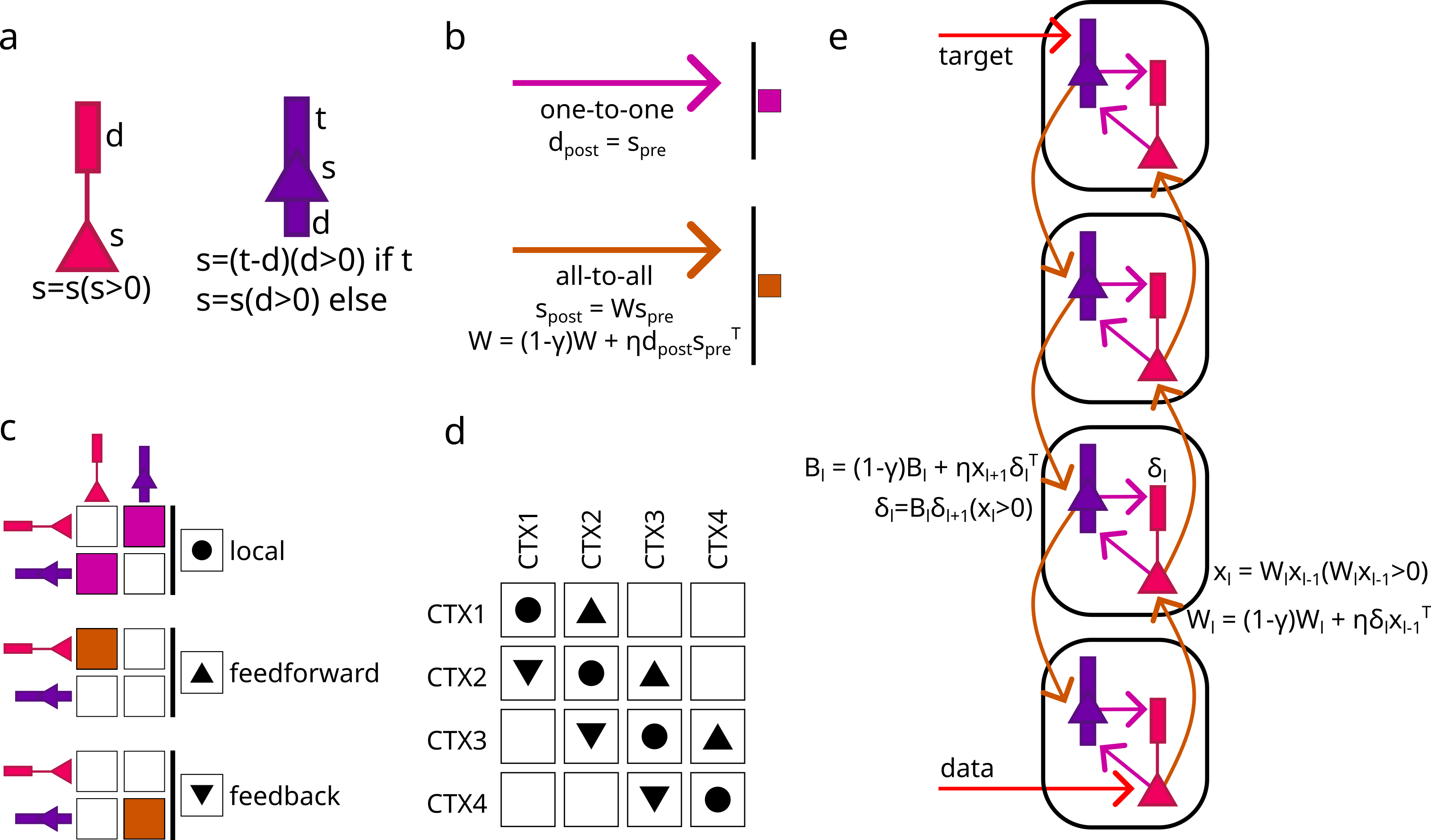}
  \caption{A circuit implementing backpropagation. More precisely, the variation of \textcite{Kolen2002-ux} avoiding weight transport, on a multilayer perceptron with ReLU nonlinearities. (a) Two cell types using neuron models with two (red) and three (purple) compartments. s=somatic activity, d=dendritic activity, t=target. (b) Two projection types: a one-to-one non-plastic projection setting postsynaptic dendritic activity (pink) and an all-to-all projection type with learnable parameter W. (c) Canonical cell type connectivity. (d) Interareal connectivity organized in a strict hierarchy. (e) A sketch of the resulting network. } \label{fig:bp}
\end{figure}

We next give a specific example of a functional model defined within this framework. We first define two pyramidal cell types with two or three compartments, respectively named neuron type 1 (\textit{N1}) and 2 (\textit{N2}). The somatic dynamics of \textit{N1} are not influenced by its dendritic compartment, while the dynamics of \textit{N2} are, see \cref{fig:bp}a. Additionally, if the third compartment of \textit{N2} is contacted, its somatic activity will reflect the difference between the two dendritic compartments. Next, we define two projection types: a one-to-one projection setting dendritic activity ($\rightarrow$) and an all-to-all projection incrementing somatic voltage and learning with the product of the postsynaptic dendritic activity with the presynaptic somatic rate ($\Rightarrow$), see \cref{fig:bp}b. We then define three connectivity blueprints: local (\textit{N1} $\rightarrow$ \textit{N2}, \textit{N2} $\rightarrow$ \textit{N1}), feedforward (\textit{N1} $\Rightarrow$ \textit{N1}), and feedback (\textit{N2} $\Rightarrow$ \textit{N2}), see \cref{fig:bp}c. Finally, we fully characterize area-level connectivity as strictly hierarchical, see \cref{fig:bp}d. To simulate the model, we run all areas sequentially, first feedforward and then feedback, running the dynamics of all cells, local projections, and interareal projections along the direction of the flow of computation, and finally applying learning to all projections. This model (cell and projection types, cell type connectivity, areal connectivity, simulation procedure) implements the backpropagation of error algorithm, more precisely the variation of \textcite{Kolen2002-ux} avoiding weight transport, on a multilayer perceptron with ReLU nonlinearities, see \cref{fig:bp}e. Alternative examples implementing e.g.~predictive coding, target propagation, generalized latent equilibrium, etc. can be constructed following the same type of procedure. Providing targets to the third compartment of \textit{N2} neurons of the top area places the network in the classical supervised learning settings. Limitations of the assumptions taken in this example are apparent (one-to-one non-plastic synapses, strict hierarchy, negative firing rates of neurons representing the gradients, unrealistic instantaneous neuronal dynamics, underestimation of the number of clearly distinct pyramidal types, no interneurons, etc.). Nevertheless, the example shows how to construct a functional and mechanistic model out of connectivity data. Future work should answer the question: \textit{What is the algorithm implemented by a model taking more realistic assumptions at every step of the modeling process outlined here?}

\section{Canonical cortical circuit search}

\textcite{dobzhansky2013nothing} famously said that ``Nothing in biology makes sense except in the light of Evolution''.
`Making sense' of neuroscientific knowledge would then require, at some level, an analysis through the lens of evolutionary biology.
Of particular interest is the genetic control of neural circuit assembly (e.g.~\cite{barabasi2020genetic,kovacs2020uncovering,kurmangaliyev2020transcriptional,arnatkeviciute2021genome}) and how it changes on evolutionary timescales (e.g.~\cite{cisek2019resynthesizing,cisek2022neuroscience,roberts2022evolution}). 
Circuits assembled following genetic instructions endow individuals from at least some species with innate capabilities before any learning by synaptic plasticity \parencite{barabasi2024functional,barabasi2025three} and provide strong inductive biases for rapid learning \parencite{zador2019critique}. 

A fruitful although marginal line of work in machine learning is also taking inspiration from evolution, often encapsulated in the term of `neuroevolution' \parencite{floreano2008neuroevolution,baldominos2020automated,stanley2019designing,miikkulainen2025neuroevolution}.
Approaches under this umbrella term vary greatly, but are in general concerned with weight compression and/or finding good network properties through population-based, derivative-free, blackbox optimization algorithms or evolutionary strategies.
These approaches are enabled by the development of efficient such algorithms, for example, the (closely related; see \cite{akimoto2010bidirectional}) Covariance Matrix Adaptation \parencite{hansen2001completely,hansen2016cma} and Natural \parencite{wierstra2014natural} Evolutionary Strategies.
A longstanding application has been the automatic exploration of the space of neural network architectures \parencite{white2023neural}. 
Of particular interest here will be methods with indirect encoding, that is, where the genome encodes a compressed representation of the network that is then fed to a (generative) development process, in contrast to a direct encoding of each individual connection (e.g.~\cite{stanley2009hypercube,risi2010indirectly}).
Some lines of work aim to harness the power of end-to-end backpropagation for efficient weight compression (through a `genomic bottleneck'). For example, \textcite{shuvaev2024encoding} train a small (`genomic') network to reproduce weights of a big (`phenotype') network found by typical backpropagation from `barcoded' representations of pairs of neurons in the big network, while \textcite{barabasi2023complex} restrict the form of typical weight matrices ($n\times n$) to go through a small bottleneck ($(n\times g) \cdot (g \times g) \cdot (g \times n)~;~g < n$) interpreted as generating full weight matrices from rules of interactions (encoded in the $g\times g$ matrix) between a few ($g$) genes.
Closest to what we will propose, \textcite{stockl2022structure} show that applying evolutionary strategies to the space of local connectivity probabilities between different cell types leads to the emergence of innate computational primitives and motor control capability in spiking neural networks.
We aim in a similar direction, but additionally propose to include not only local but also interareal connectivity blueprints between the different cell types, and include refinement through synaptic plasticity after an initial circuit assembly.

\begin{figure}[!ht]
  \centering
  \includegraphics{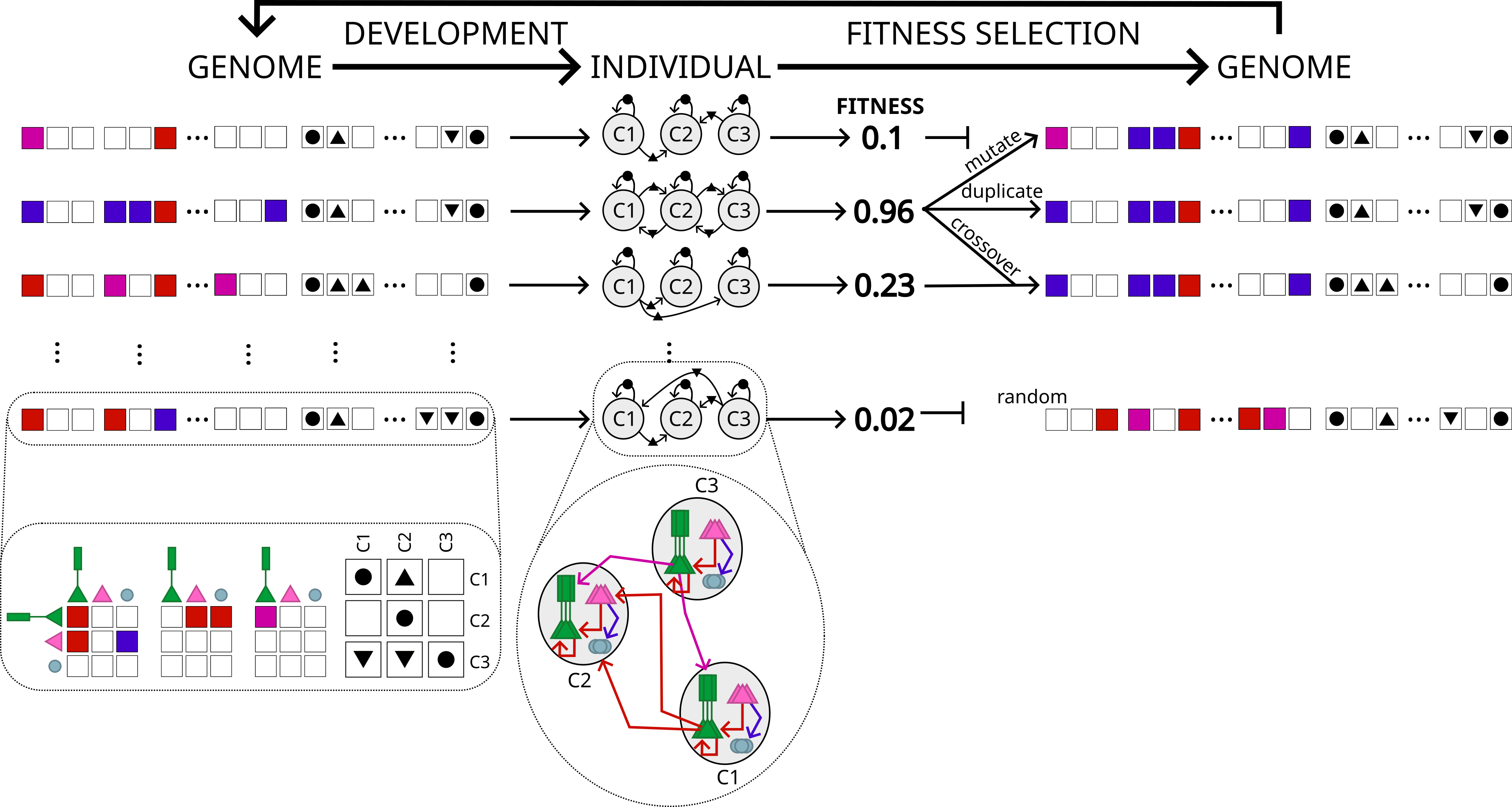}
  \caption{Canonical cortical circuit search.
  Given neuron and projection models, cell type connectivity and interareal connectivity matrices encode all the information necessary to construct actual networks.
  Here, cell type connectivity and interareal connectivity matrices are equivalently represented flattened and concatenated.
  Sets of connectivity matrices can be seen as genomes, actual networks as individuals, and the process that transforms the former into the latter as development.
  These networks can be trained and tested on functional tasks, and the underlying genomes filtered based on functional performance.
  This is akin to natural selection.
  Evolutionary search methods would then build a new set of genomes by applying crossover and mutation and repeat again the same steps.} 
  \label{fig:evolve}
\end{figure}

Here we describe the use of blackbox optimization methods or evolutionary strategies to fill connectivity matrices (`blueprints') between cell types (\cref{fig:general}c) such that the resulting architecture becomes functional.
That is, a form of neural architecture search where the search space is a subspace of the space of models defined above.
This constitutes a potentially complementary approach to the mapping of those matrices through experiments and analysis of actual cortical data.
With this approach, a network is first built by following the rules of connectivity described by the connectivity matrices, see \cref{fig:evolve} (development), and simulated with the dynamics and learning rules described by the single neuron and projection models.
When describing the resulting architecture as functional, we mean that such network is able to solve a given task, see \cref{fig:evolve} (fitness selection).
However, explicitly computing how to change terms of connectivity matrices based on the resulting network performance (as the gradient of a loss through development) appears difficult.
Hence, gradient-free optimization or search methods, and in particular evolutionary algorithms, are of particular interest, see \cref{fig:evolve}. 
The genome ($\sim\!10^9$ bits) too must encode a compact plan for building the cortex ($\sim\!10^{14}$ synapses), which then unfolds through precise developmental mechanisms. Our proposed representation of cortical connectivity is indeed highly compact, making full use of the redundancy in cortical wiring. For example, let us take a model with $a=10$ areas, $t=10$ cell types, populations of $n=100$ neurons per types per areas, $s=2$ single synapse models, $p=4$ projection types, and $b=4$ connectivity blueprints. Encoding a dense connectivity matrix of single cells takes $\log_2(s)a^2t^2n^2=10^{10}$ bits. The focus on connectivity among cell types rather than single cells bypasses the scaling in the number of single cells, lowering the encoding to $\log_2(p)a^2t^2=2\cdot10^5$ bits. Furthermore, the key assumption that the connectivity between two cortical areas necessarily follows one of b cell type connectivity blueprints greatly reduces the scaling with the number of areas, down to $\log_2(p)bt^2+\log_2(b)a^2=2000$ bits. The encoding of connectivity could even potentially be reduced further into a few rules of connectivity applied to a combinatorial code specifying cell type identity \parencite{haber2023structure,harth2024dissecting}; we do not explore this option here.

\begin{figure}[!ht]
  \centering
  \includegraphics{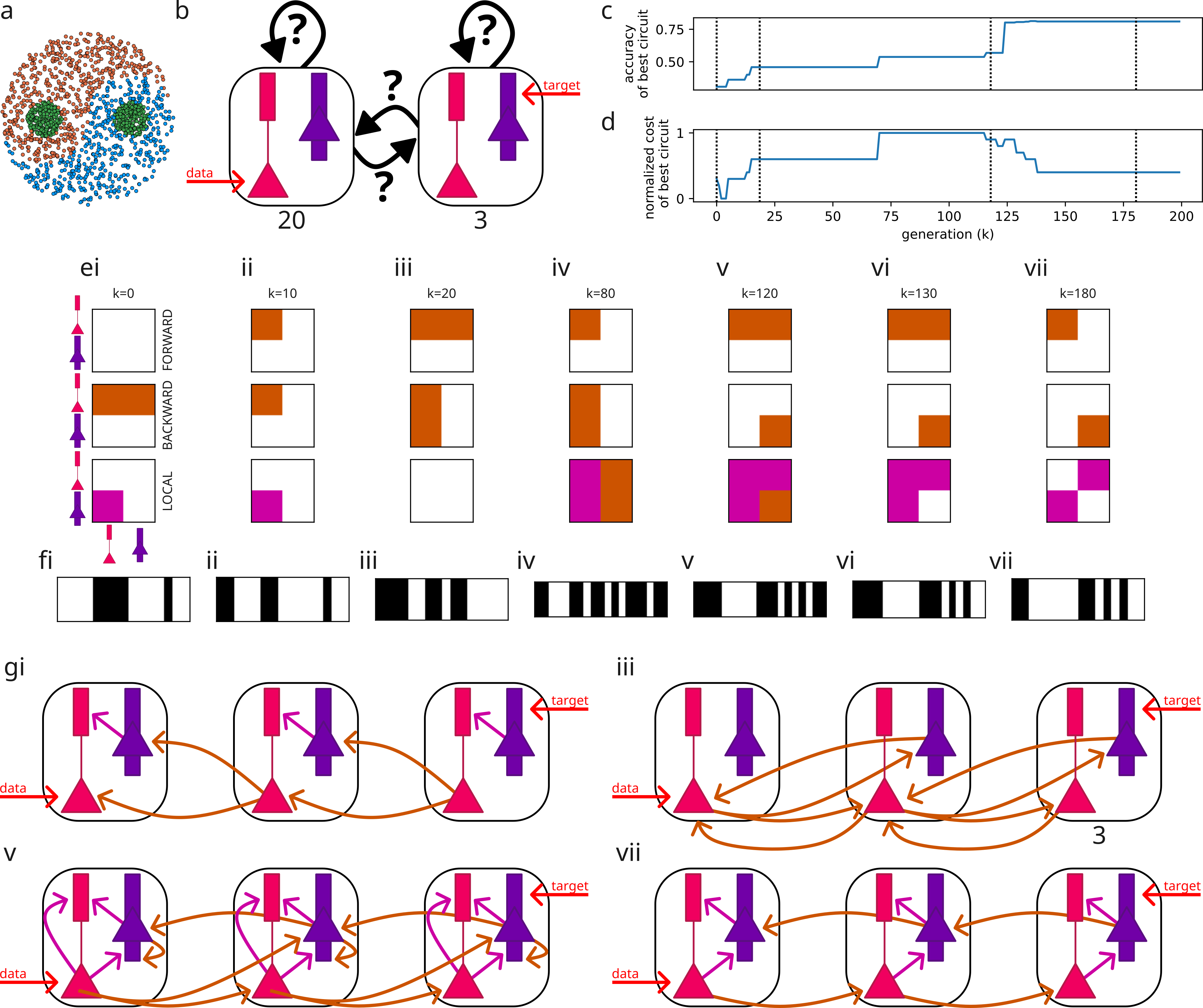}
  \caption{Finding a circuitry for backpropagation through search. 
  (a) The YinYang dataset \parencite{Kriener2021-vd}. Inputs are the $(x,y)$ coordinates; target labels are represented by the colors. 
  (b) A sketch of a small network with one hidden layer comprising 20 units. Arrows marked with `?' indicate existing connections wih unknown (varying) circuitry (cell-type-specific connectivity). 
  (c) Accuracy of the best found circuitry (applied to a network as depicted in panel b) during the search process, that is, across different generations (populations with varying circuitries). Vertical lines correspond to the generations for which the best found circuits are sketched in panel g.
  (d) Similar to panel c but reporting the normalized `metabolic' cost of the best found circuitry across generations. This cost scales with the number of existing projections between cell-type-specific populations.
  (e) Circuitries or canonical cell type connectivities with best performances in the population for different generations during the search process.
  (f) Compressed binary encodings of the cell type connectivity matrices reported in panel e.
  (g) Sketches of circuitries for corresponding subpanels i, iii, v, vii of panel e. } 
  \label{fig:evosim}
\end{figure}

As a proof of concept, we report that the circuitry for backpropagation presented in the previous section is found through evolutionary search for the best canonical circuit with two cell types and two projection types on a small supervised classification dataset (\cref{fig:evosim}a; \cite{Kriener2021-vd}). Specifically, we search the space of cell type connectivity matrices, fixing the two neurons and two projections as previously described for backpropagation (\cref{fig:bp}ab), a strictly hierarchical organization of areas, and a simulation procedure with forward and backward phases. We further reduce the search space by assuming that one-to-one synapses setting dendritic activity ($\rightarrow$) can only occur locally, and keep track of the fitness of already tested architectures in memory to accelerate computation. We simulate populations of $32$ small hierarchical networks (\cref{fig:evosim}b; one hidden layer) on $10^4$ training examples with different cell-type-specific connectivities. We employ the simplest evolutionary algorithm, with one random mutation and conservation of the best genome at every generation (elitism), and no crossover (the specifics of the algorithm do not matter for the point we make here). In other words, we try `plausible' ways of wiring cell types in local, feedforward, and feedback connections. We report the accuracy of the best resulting circuitries over generations (\cref{fig:evosim}c). To break the tie between functionally equivalent circuitries, we impose a small preference for circuitries with less connections, interpreted as minimizing metabolic cost (\cref{fig:evosim}d) for circuitries with equivalent accuracies. We report the best circuitries for different generations (\cref{fig:evosim}e), their compressed binary encodings (\cref{fig:evosim}f), as well as sketches for corresponding circuitries (\cref{fig:evosim}g). Note that at the latest stages of evolutionary search, the best found circuitry is the one that implement backpropagation/Kolen-Polack learning as described in the previous section (\cref{fig:evosim}evii, gvii; \cref{fig:bp}). For completness, we test the best found circuitry in a bigger network with two hidden layers on $10^6$ training examples, and report that this network reaches 96\% accuracy, on par with a classical multilayer perceptron trained through backpropagation \parencite{Kriener2021-vd}. The claim here is that, if we had the single neuron and projection models and the interareal connectivity, but did not know the backpropagation algorithm, we could have found (a neural circuit that implements) it through evolutionary search. 

A similar process is in theory applicable to the as yet unknown canonical cortical algorithm. In principle, then, good single neuron and projection models are all we need, in addition to the compact representation of connectivity rendering the search computationally feasible. In practice, the search space remains large and a further reduction of acceptable models by known properties of cortical structure is warranted. Another important practical aspect is the design of the benchmark or curriculum on which the different architectures are tested. Starting with simple machine learning tasks that should quickly differentiate functional from unfunctional architecture seems crucial \parencite{Kriener2021-vd}. Another potential target is the generation of stable, balanced, biologically plausible dynamics \parencite{Kurth2024-se}.

\section{Discussion}

In this work, we review recent findings regarding the functional response patterns of cortical cell types, and conclude that cell types fulfill distinct computational roles. We also propose a perspective on the current knowledge of cortical structure, and conclude that it can be efficiently recapitulated through canonical connectivity blueprints between cell types. We outline a general framework to build both functional and mechanistically scrutable models with these two principles as a basis, and propose to search the space of such models for functional architectures. A similar focus on the computational specialization of different cell types and their connectivity has been successfully applied to understand the computation performed by neural circuits of other brain structures, notably the retina \parencite{Seung2014-qw}, the fly central complex \parencite{Turner-Evans2020-bh,Hulse2021-ob}, and, more recently, the songbird forebrain \parencite{Hozhabri2025-bn}.

This bottom-up approach stands in contrast to a more top-down, normative approach, where circuits are built to implement theoretically-derived neural dynamics \parencite{Ellenberger2024-iv,Senn2024-ew,George2025-tc}. Moreover, our desire for a mechanistic understanding leads us to focus on neural circuits, rather than on other, related levels of abstraction, such as neural manifolds and dynamics \parencite{Langdon2023-ji,Ostojic2024-gq}. In this work, we focus on structural connectivity. Models derived in our framework necessarilly depend on single neuron and projection models, and these are likely as crucial to our understanding as connectivity (\cite{Bargmann2013-ra}; but see \cite{Lappalainen2024-ut}). The modelling of these basic building blocks should also be constrained by both experimental and theoretical studies \parencite{Eckmann2024-mu,Moore2024-ke,Moldwin2025-mg}. For example, with the linear relationship between synapse size and strength \parencite{Holler2021-wb}, a distribution of synaptic weights for synapses connecting two specific cell types could be obtained from electron microscopy data (e.g., \cite{Sievers2024-nv,microns2025functional}), and used to constrain projection models.

We have focused on cortex, but a more complete understanding of cortical computation will certainly include the ubiquitous thalamocortical interactions \parencite{Suzuki2023-hg,Mo2024-gv,Wolff2024-sg,mckinnon2025disruption}. A coherent picture of the function of corticothalamic interactions, and notably of `transthalamic' pathways connecting two cortical areas through the thalamus, has yet to emerge. As a potential first step, \textcite{Sherman2024-aw} propose to consider corticothalamic projections originating from pyramidal cells of layer 5 as transporting efference copies of motor commands. Indeed, these projections typically branch to subcortical motor centers, and every cortical area sends such projections, including primary sensory areas. Based on the structure of cortico-thalamic circuits, \textcite{granier2025multihead} propose an analogy to multihead self-attention, the central algorithmic motif in transformer networks. Perhaps, another clue lies in the observed parallels in the structure and computation of thalamus and cortical layer 6b \parencite{Zolnik2024-is}. Our modelling framework very naturally extends to the definition of other brain structures beyond cortex and their cell type connectivity, both intra- and inter- structures.

Future work should investigate different, likely less discrete indirect encoding of connectivity, notably in the context of neuroevoution. This includes the encoding of pre- and post-cell-type-specific connection probability and/or distance rule in $[0,1]^{t^2}$, with $t$ the number of cell types. We will additionally investigate the further compressing of connectivity blueprints between cell types, notably taking a form similar to the one of \textcite{barabasi2020genetic} with a genetic bottleneck ($(t\times g)\cdot(g\times g)\cdot(g\times t), g<t$), whereby the connectivity between cell \textit{types} (instead of single cells) can be interpreted as resulting from the rules of interactions between a few genes. Parameters of neuronal models (notably time constants, but also parameters of simple neuron models, e.g.~the 4 parameters of the model of \textcite{izhikevich2003simple}) and learning rules \parencite{jordan2021evolving,confavreux2023meta,confavreux2025balancing,confavreux2025memory} and interareal connectivity (potentially again a compressed representation, down perhaps to a one-parameter rule based on the distance between two areas; \cite{ercsey2013predictive}) might be added to the search space. Finally, we point to a curriculum of tasks aiming at adaptive complex motor control as a particularly interesting way to define fitness.

\section{Acknowledgments}
We thank Anna Vasilevskaya, Roman Doronin, and Georg Keller for helpful discussions of the reviewed data and feedback on the structure of the manuscript.
This work was funded by the European Union’s Horizon Europe Programme under the Specific Grant Agreement No.~101147319 (EBRAINS 2.0 Project).

\section{Code Availability}
All code written in support of this publication is publicly available at \url{https://github.com/arneuro/neuroevo-bp}. 

\printbibliography
\end{document}